\documentclass[apj]{emulateapj}
\def\simgt{\lower.5ex\hbox{$\; \buildrel > \over \sim \;$}}
\def\simlt{\lower.5ex\hbox{$\; \buildrel < \over \sim \;$}}
\def\sp{\hspace{1.5pt}}

\def\sp{\hspace{1.5pt}}

\def\amin{\ifmmode^{\prime}\else$^{\prime}$\fi}
\def\asec{\ifmmode^{\prime\prime}\else$^{\prime\prime}$\fi}

\def\simgt{\lower.5ex\hbox{$\; \buildrel > \over \sim \;$}}
\def\simlt{\lower.5ex\hbox{$\; \buildrel < \over \sim \;$}}

\newcommand\chandra{{\it Chandra}}

\newcommand\xmm{{\it XMM\/}-Newton}
\newcommand\XMM{{\it XMM\/}-Newton}

\def\sp{\hskip 1.5pt}
\def\psr{\hbox{PSR~J0205+6449}}
\def\kes73{\hbox{Kes\sp73}}

\shorttitle{Thermal X-ray Shell Around SNR 3C58}
\shortauthors{Gotthelf et al.}

\begin{document}

\title{A Shell of Thermal X-ray Emission Surrounding the Young
 Crab-like Remnant 3C58} \author{E. V. Gotthelf, D. J. Helfand, \&
 L. Newburgh}

\affil{Columbia Astrophysics Laboratory, Columbia University, 550 West 120$^{th}$ Street, New York, NY 10027, USA; eric@astro.columbia.edu}





\begin{abstract}

Deep X-ray imaging spectroscopy of the bright pulsar wind nebula 3C58
confirms the existence of an embedded thermal X-ray shell surrounding
the pulsar \psr. Radially resolved spectra obtained with the \xmm\
telescope are well-characterized by a power-law model with the
addition of a soft thermal emission component in varying proportions.
These fits reproduce the well-studied increase in the spectral index
with radius attributed to synchrotron burn-off of high energy
electrons. Most interestingly, a radially resolved thermal component
is shown to map out a shell-like structure $\approx 6^{\prime}$ in
diameter. The presence of a strong emission line corresponding to the
\ion{Ne}{9} He-like transition requires an overabundance of $\sim 3 \times
[{\rm Ne/Ne}_{\sun}]$ in the Raymond-Smith plasma model. The best-fit
temperature $kT \sim 0.23$ keV is essentially independent of radius
for the derived column density of $N_{\rm H} = (4.2\pm 0.1) \times
10^{21}\ {\rm cm}^{-2}$. Our result suggests that thermal shells can
be obscured in the early evolution of a supernova remnant by
non-thermal pulsar wind nebulae emission; the luminosity of the 3C58
shell is more than an order of magnitude below the upper limit on a
similar shell in the Crab Nebula. We find the shell centroid to be
offset from the pulsar location. If this neutron star has a velocity
similar to that of the Crab pulsar, we derive an age of 3700 yr and a
velocity vector aligned with the long axis of the PWN. The shell
parameters and pulsar offset add to the accumulating evidence that
3C58 is not the remnant of the supernova of CE 1181.

\end{abstract}

\clearpage

\section{Introduction}

A long standing puzzle in supernova physics is the apparent lack of an
associated thermal shell surrounding a few young pulsars with bright
relativistic wind-powered nebulae (PWNe). Though relatively rare, with
fewer than ten examples known, the lack of a supernova remnant shell
is at odds with our current understanding of supernova evolution and
neutron star formation. The canonical examples, the Crab and 3C58, are
both putative historical remnants less than 1000 yrs old.  To date, no
evidence has been found for thermal emission associated with the Crab
Nebula pulsar despite repeated searches (Seward et al. 2006 and
references therein). However, recent \xmm\ and \chandra\ observations have
identified thermal emission within the 3C58 remnant (Bocchino et
al. 2001; Slane et al. 2004).  The availability of a large quantity of
XMM archival data on this source derived from calibration observations
allows us to construct very sensitive images and spectra to explore
the evidence for a surrounding SNR shell with unprecedented
sensitivity.

The morphology of 3C58 is well-documented in both the X-ray and radio
bands (Slane et al. 2004 and Bietenholz et al. 2006 and references
therein). It is center-filled with a 66~msec pulsar near the center,
axisymmetric lobes, and an elaborate network of wisps and filaments.
Failure to find a radio shell around 3C58 led Reynolds \& Aller (1985)
to conclude that there is no evidence for its interaction with an
external medium.  Coupled with the low velocities of the optical
filaments (Fesen 1983) compared to the mean expansion velocity
required for the remnant to reach its current extent in 820 yr, these
results led to suggestions that the remnant was considerably
older. Over the past twenty years, further evidence has accumulated
that the remnant's age is inconsistent with an origin in SN1181. The
images and spatially resolved spectra we present below add
incrementially to the case and provides a testable prediction
concerning the pulsar velocity which could support or refute the
historical association.

\begin{deluxetable*}{lccccc}
\tabletypesize{\small}
\tablewidth{0pt}
\tablecaption{\xmm\ Observation log for 3C58$^a$}
\tablehead{
\colhead{ObsID} & \colhead{Date} & \colhead{Pointing}& \colhead{pointing} & \colhead{Exposure}       & \colhead{Exposure} \\
\colhead{     } & \colhead{}   & \colhead{R.A}   & \colhead{Decl.}    & \colhead{EPIC-MOS$^b$} & \colhead{EPIC-pn}\\
\colhead{     } & \colhead{(UT)} & \colhead{(J2000)} & \colhead{(J2000)}  & \colhead{(ks)}           & \colhead{(ks)}     \\
}
\startdata
153752201\sp & 2002-09-11~04:29:35 & $2^{\rm h} 4^{\rm m}43.6^{\rm s}$ & $+64^{\circ}51^{\prime}13{\farcs}2$ & 36777 & 15579 \\
153751801\sp & 2002-09-11~13:09:58 & $2^{\rm h} 5^{\rm m}03.0^{\rm s}$ & $+64^{\circ}47^{\prime}38{\farcs}6$ & 39561 & 16328 \\
153752501\sp & 2002-09-11~20:03:41 & $2^{\rm h} 5^{\rm m}23.4^{\rm s}$ & $+64^{\circ}43^{\prime}52{\farcs}3$ & 35195 & 14721 \\
153752401\sp & 2002-09-12~03:20:44 & $2^{\rm h} 6^{\rm m}32.4^{\rm s}$ & $+64^{\circ}48^{\prime}06{\farcs}8$ & 26184 & \dots \\
153752101    & 2002-09-13~04:22:11 & $2^{\rm h} 5^{\rm m}38.0^{\rm s}$ & $+64^{\circ}49^{\prime}40{\farcs}0$ & 17363 &  3022 \\
153751701    & 2002-09-13~10:57:39 & $2^{\rm h} 5^{\rm m}18.9^{\rm s}$ & $+64^{\circ}53^{\prime}23{\farcs}7$ & 13515 &  6042 \\
153751901\sp & 2002-09-13~17:51:22 & $2^{\rm h} 5^{\rm m}13.1^{\rm s}$ & $+64^{\circ}51^{\prime}41{\farcs}8$ & 34114 & 14569 \\
153752001\sp & 2002-09-13~23:45:05 & $2^{\rm h} 5^{\rm m}52.6^{\rm s}$ & $+64^{\circ}55^{\prime}27{\farcs}7$ & 48631 & 21038 \\
\enddata
\tablenotetext{a}{\footnotesize All observations were obtained in pointed imaging mode with the thin transmission window in the focal plane; the EMOS and EPN data were collected with the {\tt PrimeFullWindow} and {\tt PrimeLargeWindow} submodes, respectively}
\tablenotetext{b}{\footnotesize  EPIC exposure times are after filtering; the EPIC-MOS exposure times are for the sum of the two MOS cameras.}
\label{ta:log}
\end{deluxetable*}

In \S2 we describe the available observations and discuss their
analysis. Section 3 presents the detailed spatially resolved
spectroscopy which allows us to detect clear evidence for a shell of
thermal emission, overabundant in Neon, and extending to the edge of
the synchrotron nebula; we interpret this result in \S4, concluding
that the pulsar is offset from the center of the symmetrically
expanding shell implying a velocity aligned with the symmetry axis of
the remnant. If the pulsar has a Crab-like velocity, this offset
provides another argument against an association with SN1181.

\section{Observations and Results}

\subsection{The observations}

A total of ten pointed observations of 3C58 are available in the
public archive of data obtained with the \xmm\ X-ray Observatory
(Jansen et al. 2001). Of these, a set of eight pointings, acquired on
2002 September $11-13$ to perform an in-orbit calibration of the
telescope mirror vignetting function (Lumb et al. 2004), used imaging
modes with a sufficiently large field-of-view to cover fully the 3C58
nebula. The duration of these observations varied between
$17-33$~ks. An observation log is presented in Table 1.

\begin{figure*}
\centerline{
\includegraphics*[angle=270,scale=0.30]{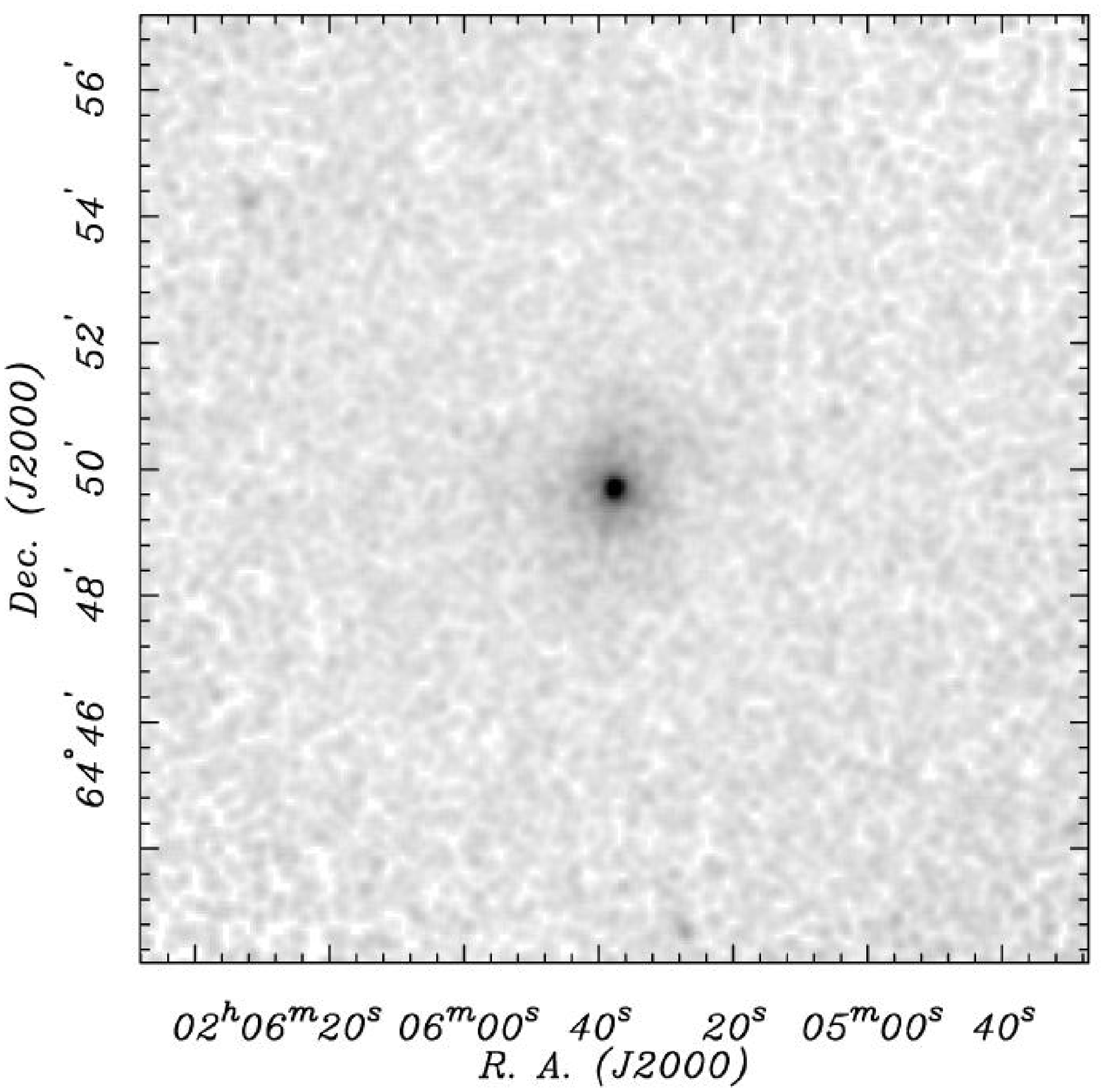}
\hfill
\includegraphics*[angle=270,scale=0.30]{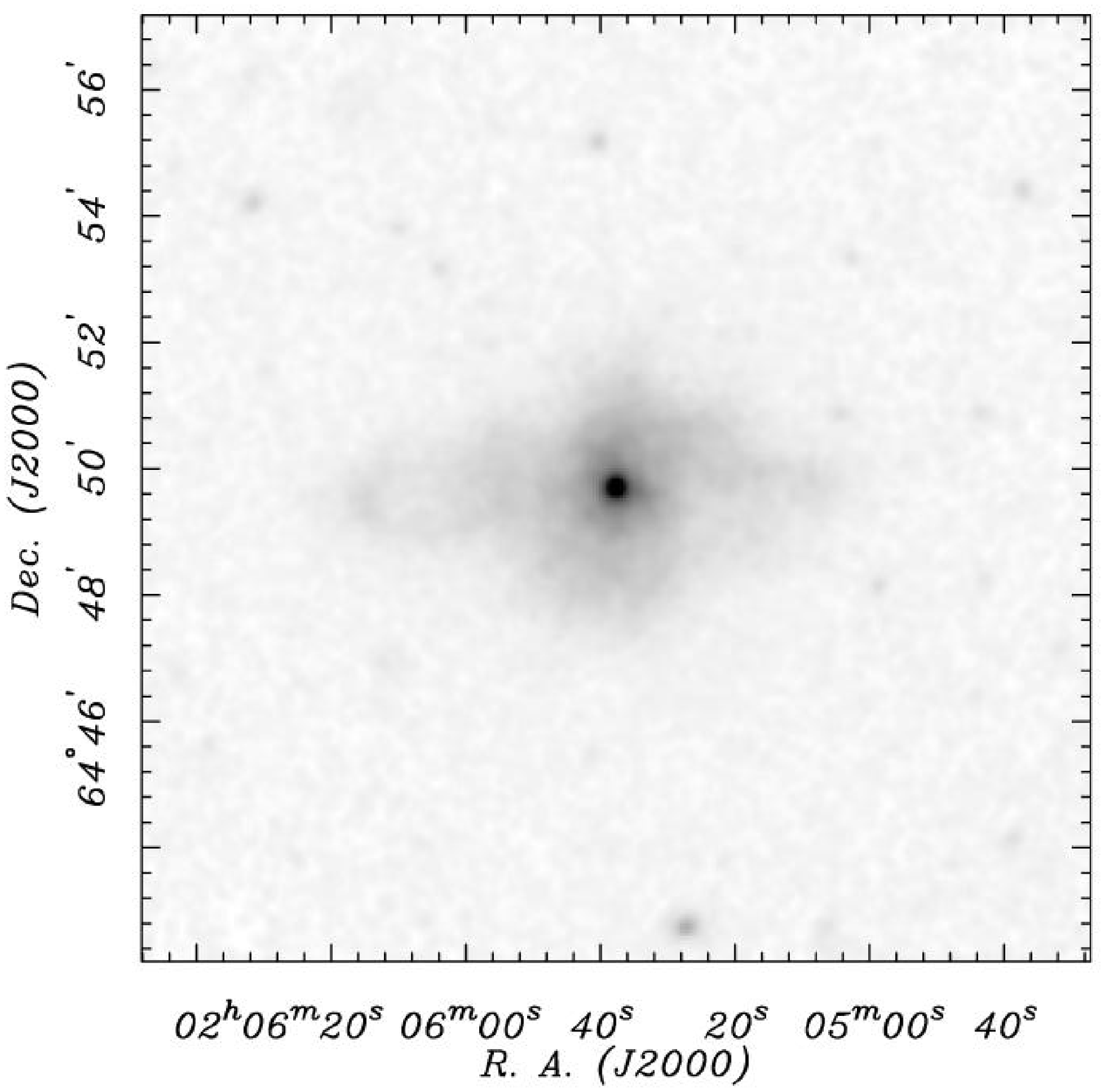}
\hfill
\includegraphics*[angle=270,scale=0.30]{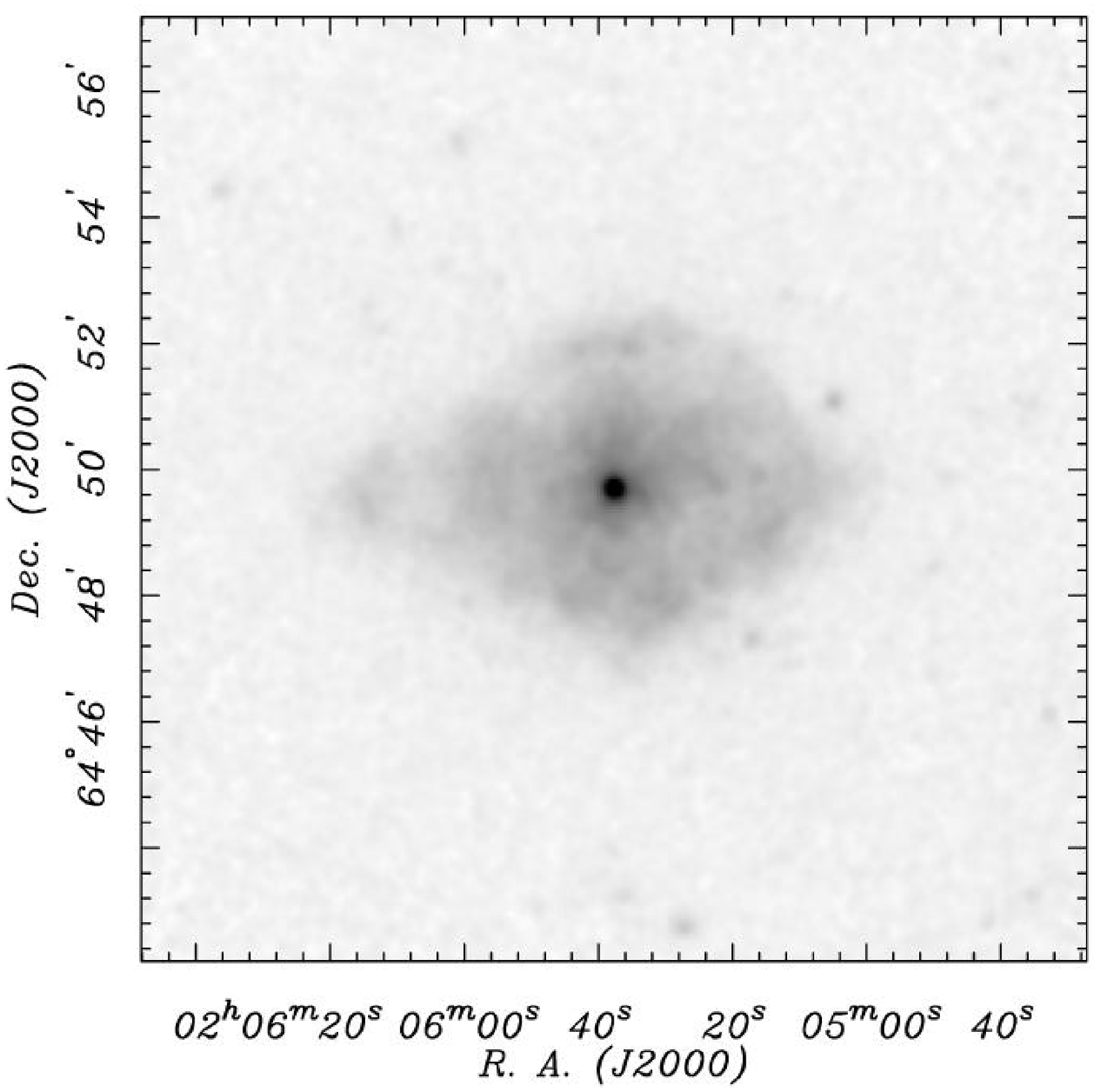}
}
\caption{Narrow-band X-ray images of 3C58 acquired with the \xmm\ EPIC
cameras. The energy range for these exposure-corrected images
separates out three distinct sources of emission, {\it Left)} the
unresolved point-like pulsar emission surrounded by the highest-energy
and most intense portion of the PWN ($7.0-8.0$~keV), {\it Middle)} the
pulsar wind nebula in the medium band ($2.0-3.0$~keV), and {\it
Right)} the symmetric shell-like structure in the soft X-ray band
($0.5-1.0$~keV). The pulsar wind nebula image also includes pulsar
emission whereas the shell-like structure includes both pulsar and PWN
emission components.  In each image, the intensity is cropped at $0.6$
of the peak intensity and plotted with square-root scaling to highlight the
fainter emission. }
\end{figure*}

\begin{figure}
\centering
\includegraphics*[angle=270,width=0.9\linewidth,clip=]{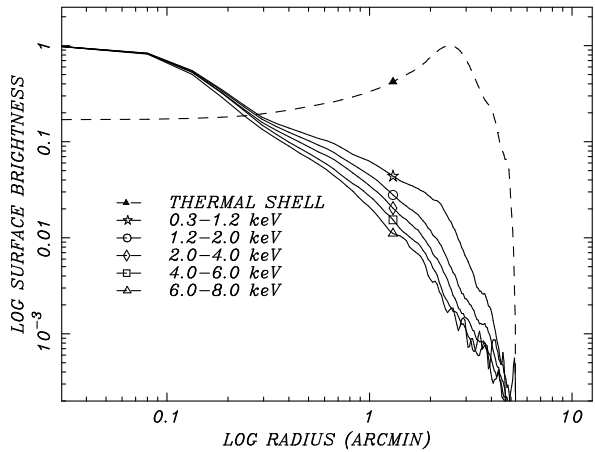}
\caption{Evidence of a thermal shell surrounding the pulsar in 3C58.
Normalized background-subtracted radial profiles for five adjacent energy
bands from $0.3-8.0$ keV obtained with the \xmm\ EPIC CCD detectors. A
prominent plateau of emission is seen for the lowest energy
profile. A shell-like structure ({\it dashed-line}) is found for the
radial profile of the $0.3-1.2$ keV band image after subtracting off
the normalized emission in the next highest band.}
\end{figure}

\begin{figure}
\centering
\includegraphics*[angle=270,width=0.9\linewidth,clip=]{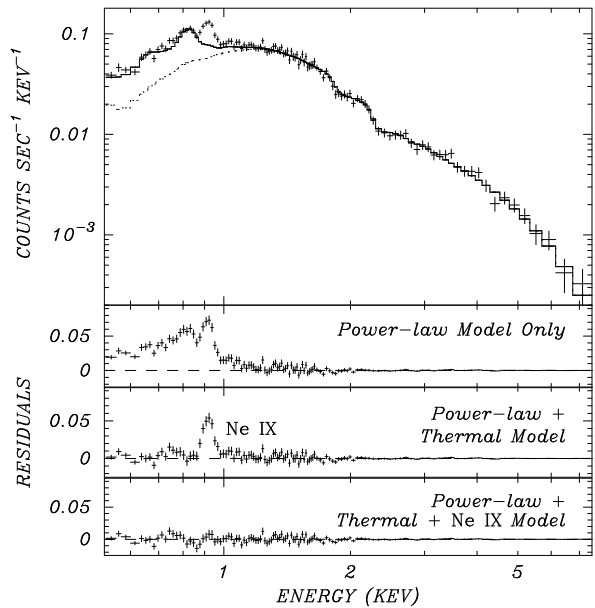}
\caption{Evidence for a thermal component with an overabundance of
Neon in 3C58. The solid line in the upper panel represents a best-fit
power-law plus Raymond-Smith plasma emission model to the \xmm\
EPIC-MOS spectra ({crosses}) from annular regions ($2\farcm0 -
2\farcm5$) centered on the pulsar in 3C58. The dotted line shows the
power-law component only. Residuals from three different fits are
displayed in the lower three panels. The first of these shows a strong
excess above the pure power-law model below 1.2~keV. The next panel
down shows the residuals from a model that includes an equilibrium
thermal R-S plasma with solar abundances. The strong line remaining at
$\sim0.9$~keV corresponds to a complex of emission lines from
Helium-like Neon (\ion{Ne}{9}). The bottom panel shows a fit in which
the Neon abundance is three times solar. Model parameters are given in
Table 2.}
\end{figure}

In this study we concentrate on data obtained with the European Photon
Imaging Camera (EPIC; Turner et al. 2001) which consists of three CCD
cameras, the EPIC-pn and the two EPIC-MOS imagers. The EPIC-MOS
cameras were operated in {\tt FullFrameMode} mode, in which the full
$30^{\prime}$ diameter field-of-view is read out every 2.7~s. The
EPIC-pn was operated in {\tt PrimeLargeWindow} mode, which provides
78~ms time resolution over a $ 6\farcm9 \times 13\farcm5$
field-of-view. For all observation, the thin transmission filter was
placed in the focal plane. The EPIC instruments are sensitive to
X-rays in the $0.2-10$ keV energy range. The point spread function
(PSF) of the mirror modules has a FWHM of $6-15^{\prime\prime}$
depending on energy and is fully oversampled by the CCD pixel size in
each instrument.

\subsection{Basic analysis}

We analyzed data produced by the standard processing (SAS version {\tt
20020507\_1701}).  The photon event lists were initially filtered for
periods of high background, typically flare events corresponding to
enhanced solar wind activity. To identify these flares, an iterative
clipping method was used to generate the acceptable time intervals
using the following method. We first produced a lightcurve histogram
in 100~s steps for the whole instrument and calculated the
mean. Intervals with an enhanced count rate $> 3\sigma$ above the mean
were discarded and a new lightcurve was generated. This process was
repeated until all points above the converging threshold were
removed. The good time intervals (GTIa) thus generated were used to
filter the revised event file. A total of $\sim 343$ ks of good
exposure time was acquired from the collective data set; the net
exposure times for each instrument are listed in Table 1. Finally, in
creating spectra and images, we used the standard SAS screening flags
and selected only CCD photons with {\tt PATTERN} $\le 4$ and $\le 12$
for the EPIC-pn and EPIC-MOS, respectively. Bad CCD pixels and columns
were taken into account.  We verified that spectra were not affected
by photon pile-up in any of the observations.

{\subsection{Spatial analysis}

To search for evidence of a faint supernova remnant surrounding the
pulsar, we generated exposure-corrected, narrow-band images including
data from all three instruments in order to obtain the deepest
possible image. Figure~1 compares the image of 3C58 in three energy
bands to highlight the three spatial components which are found to
make up the composite morphology: a shell-like nebula ($0.5-1.0$~keV),
the pulsar wind nebula ($2.0-3.0$~keV), and the point-like pulsar
emission ($7.0-8.0$~keV).  This is best illustrated by comparing the
radial profiles in the five narrow energy-bands shown in Figure~2.  In
addition to the steep falloff of flux with radius, it is again clear
that the extent of the emission from 3C58 is strongly dependent on
energy-band. In particular, the radial profile in the softest energy
band displays a large departure from the higher energy profiles
between $2^{\prime}$ and $5^{\prime}$ from the remnant
center. Subtracting emission in the $1.2-2.0$ keV band from the lowest
energy image shows that the soft emission has a radial profile
consistent with that of a limb-brightened shell peaking $\sim
3^{\prime}$ from the remnant center.

To investigate further the nature of this shell-like emission we
generated spectra from concentric annuli $30^{\prime\prime}$ wide to
search for radial variations.  These annuli were centered on the
location of the pulsar, the presumed origin of the wind and likely
center of any SNR expansion. To estimate the background in each
annulus, we collected photons from within a region with $\sim 3\farcm5
< r < 4\farcm0$. Although the background is somewhat distant from the
inner source regions, the high surface brightness of the source makes
background subtraction less important there.  We also extracted
photons from the eastern ``lobe'' region, using a circle of radius
$75^{\prime\prime}$ centered at coordinates RA = 02:06:14.65 and Dec =
64:49:31.9; the background for this region was determined from a pair
of apertures straddling the source region to the north and south.

\subsection{Spectral analysis}

Spectra from each observation were extracted from each annular region
using standard channel binning for each instrument, and a set of
response matrices were generate for each ring using the standard
prescription for diffuse emission. The spectra and response matrices
were then summed over all observations for each instrument following
the method used in {\tt addascaspec}\footnote{{\tt addascaspec} is
part of the FTOOLS software package available at {\tt
heasarc.gsfc.nasa.gov/docs/software.html}}. The two sets of MOS
spectra were then combined. Finally, all spectra were grouped into
bins containing a minimum of~400 counts and fitted using the {\tt
xspec} spectral fitting package.

\begin{figure*}
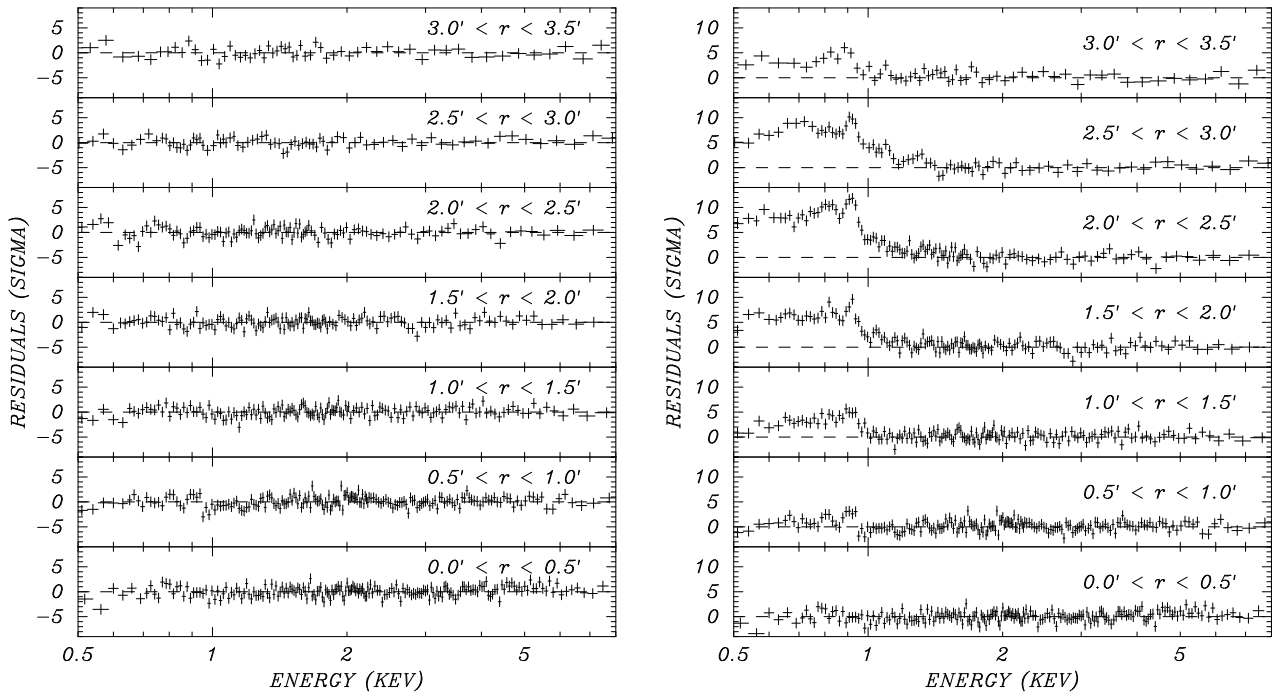

\centerline{
\hfill
 \includegraphics*[angle=270,width=0.45\linewidth]{f4a.eps}
\hfill
 \includegraphics*[angle=270,width=0.45\linewidth]{f4b.eps}
\hfill
}
\caption{ {\it Left Panel: } Residual from the best fit to the \xmm\
EPIC-MOS spectrum of 3C58 in units of sigma as a function of radius.
Data are accumulated in concentric annuli centered on the pulsar as
defined in Table~2 and are ordered by decreasing radius (top to
bottom).  Each plot shows data from the combined EPIC-MOS spectrum
from eight observations of the pulsar; a similar result is found for
the EPIC-PN camera (not shown). The data in all cases are well-fit by
a two-component power-law plus Raymond-Smith model with the parameters
given in Table~2.  {\it Right Panel: } As for adjacent panel, but with
the thermal component normalization set to zero to show this
contribution as a function of radius. The flux of the thermal
component as a function of radius is also displayed in Figure~6.}
\end{figure*}

\begin{figure}[!t]
\centerline{
 \includegraphics*[angle=270,width=0.9\linewidth]{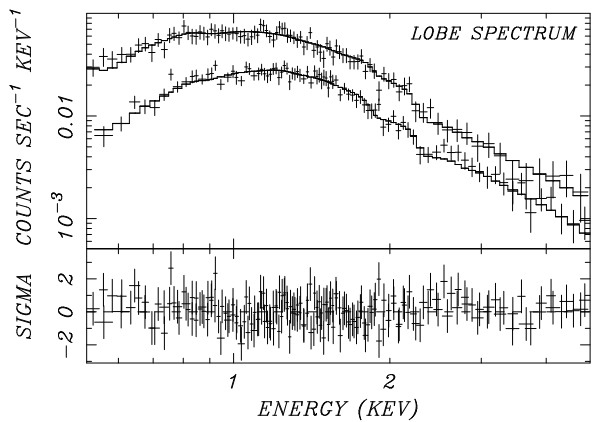}
}
\caption{\xmm\ EPIC spectrum of the 3C58 lobe region (see text) fitted to a
power-law model. The bottom panel displays the residuals from the best
fit model with $\Gamma = 2.88 \pm 0.05$~keV.  }
\end{figure}

\begin{figure}
\centerline{
\includegraphics*[angle=270,width=0.9\linewidth]{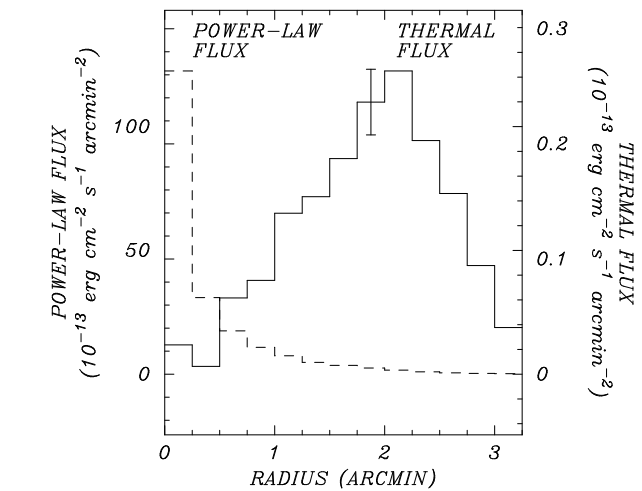}
}
\caption{Spectroscopic detection of a thermal shell surrounding
3C58. Shown are the normalized radial profiles of the component flux
from a power-law plus variable abundance Raymond-Smith plasma emission
model. The observed fluxes are given in the 0.5--10.0 keV energy band
using the EPIC-MOS cameras on-board \XMM; a similar result is found
using the EPIC-pn cameras (see Table 2). Notice the change of scale by
a factor of $\sim 500$ between the ordinate axes; less than $4\%$ of the
observed remnant flux is in the thermal component, explaining why it
has proven so illusive.}
\end{figure}

\begin{figure}
\begin{center}
\includegraphics*[angle=270,width=0.9\linewidth,clip=true]{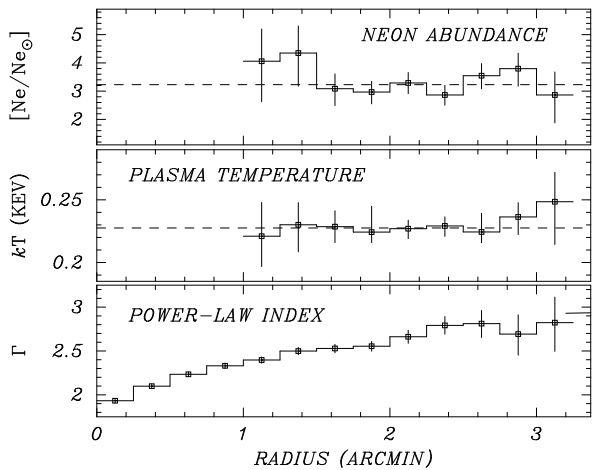}
\end{center}
\caption{Photon index for the best fit power-law model, Raymond-Smith
plasma temperature, and Ne abundance as a function of radius. The
power-law index increases monotonically from $\Gamma = 1.9$ to $\Gamma
= 2.8$, consistent with previous reports in Torii et al. (2000) and Slane et al.
(2004). The index derived from emission in the eastern lobe up to $4^{\prime}$ 
from the center is shown on the right axis. Both the plasma temerature and the 
Neon abundance show no significant radial dependence between $1\farcm0$ and 
$3\farcm2$ }

\end{figure}

\begin{figure*}
\centerline{
\hfill
\includegraphics*[angle=0,width=0.43\linewidth]{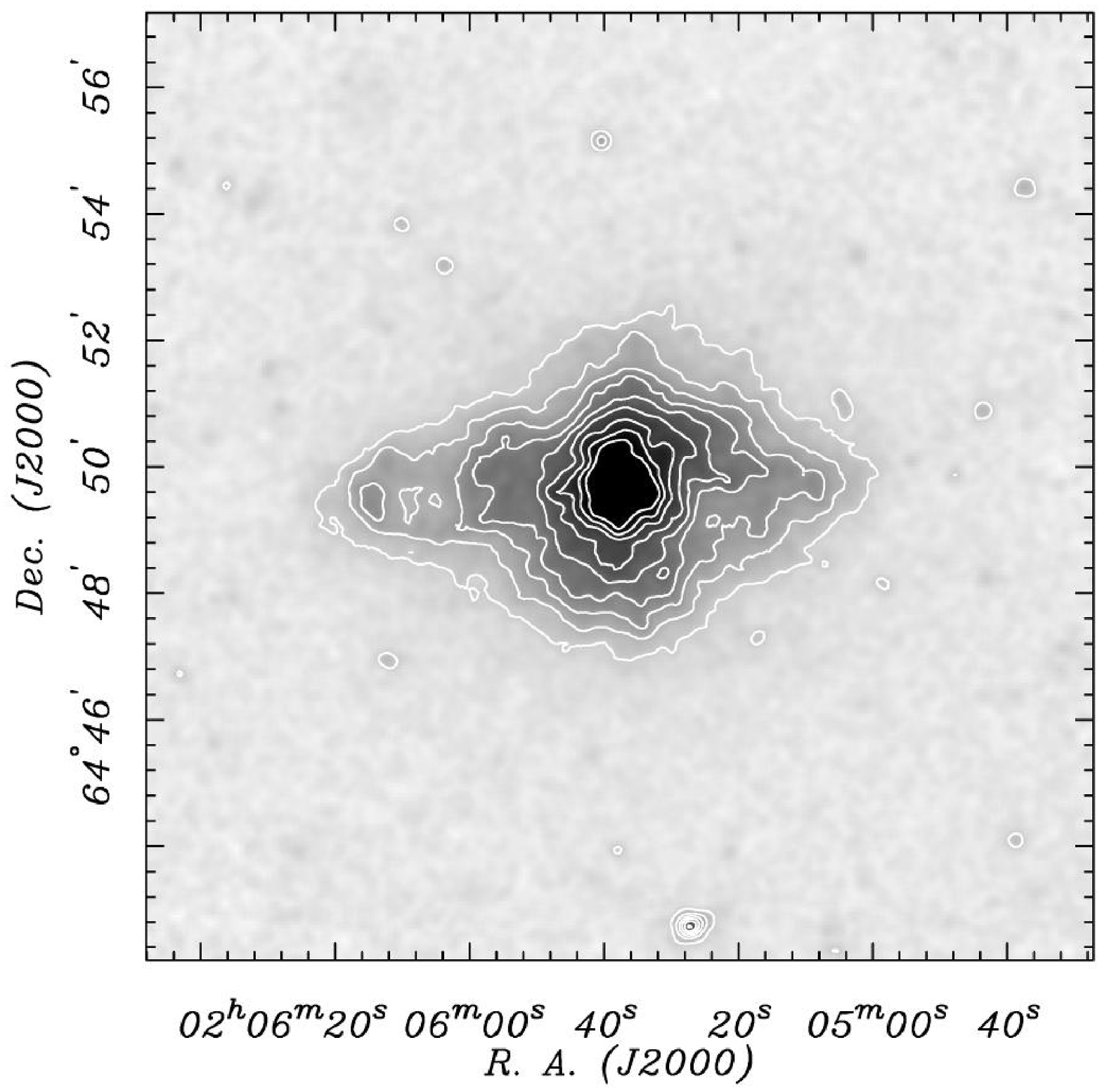}
\hfill
\includegraphics*[angle=0,width=0.43\linewidth]{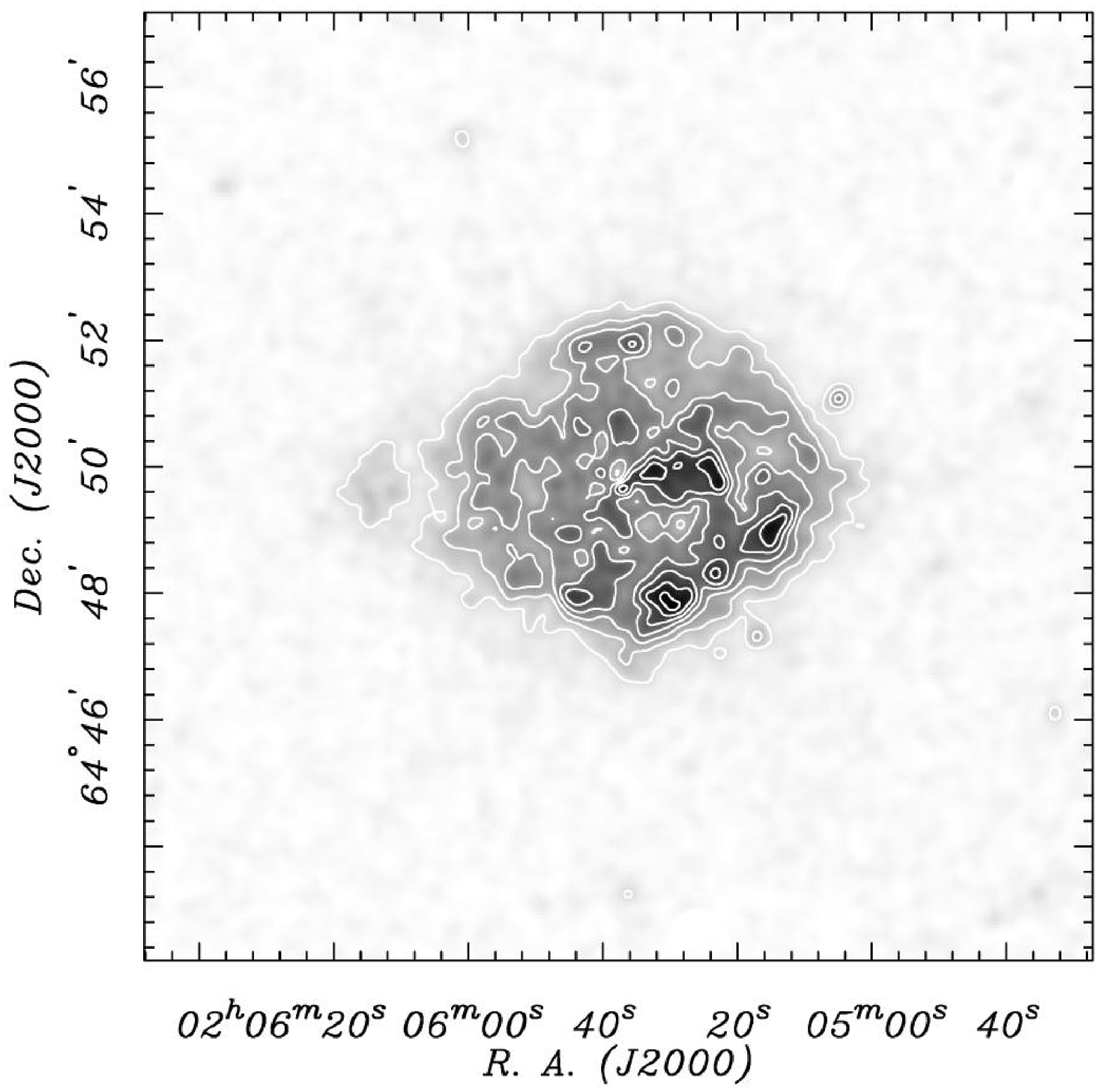}
\hfill
}
\caption{X-ray image of the pulsar wind nebula and the thermal shell
in the supernova remnant 3C58. These images are the mosaic of~8
observations acquired with the \xmm\ EPIC cameras.  {\it Left)} the
$1.0-2.0$~keV exposure-corrected image with contours illustrates the
morphology of the pulsar wind nebula and ``lobe'' regions. This image
is scaled to enhance the fainter nebula emission with contours
overlayed. {\it Right)} The same data but in the $0.5-1.0$~keV energy
band with the non-thermal contribution subtracted (see text). The
intensity is scaled linearly and the seven contours are equally
spaced. The symmetric, filled shell-like structure is $\sim 5\farcm6$
in diameter.
}
\end{figure*}

To fit the summed spectra in each annulus we used the following
procedure. In all cases, the EPIC-pn and EPIC-MOS spectra were fitted
simultaneously, with the model normalizations allowed to be
independent to account for remaining flux calibration differences
($\simlt 10\%$) between the two instruments (Saxton 2003). The spectra
were initially fitted in the $1.5-8.0$~keV energy band with a
power-law model for the non-thermal emission from the pulsar wind
nebula.  This model included the effects of interstellar absorption,
although this is not an important component in the fitted band.  The
best-fit values for the power-law index and flux are recorded in
Table~2. As illustrated in Figure~3, the spectrum above 1.2~keV is
well-characterized by a power-law model. However, at the lower
energies, we require additional emission components for the larger
annuli. We account for this excess by adding in a Raymond-Smith
thermal plasma emission component to the model and extending the
fitting range to $0.5-8.0$~keV. In fitting this new model, the
power-law indices were fixed to the values derived at high
energies. As shown in the third panel of Figure~3, however, a strong
line feature at $\sim 0.9$~keV remained unmodeled. This emission line
corresponds to the He-like \ion{Ne}{9} transition and requires an
overabundance $\sim 3 \times [{\rm Ne/Ne}_{\sun}]$ in the
Raymond-Smith plasma model throughout the nebula. 

Our final model includes a Gaussian line component to allow for the
Neon emission contribution. For these fits the N$_H$ was fixed to the
value derived from the central annulus, the region with the most
counts and in which the non-thermal emission strongly dominates over
any thermal emission.  In all cases, the final spectral fits produce
an acceptable fit statistic, with the exception of the region between
$r=2-2.5'$, and perhaps the next interior one. This can be attributed
to unmodeled structure below 0.8~keV, likely additional weak thermal
line emission which we lack the sensitivity to model with the current
data. Table~2 gives the best fit values for the temperature and flux
in each annulus after fixing the $\Gamma$ and N$_H$, while Figures~4
demonstrates the dramatic change in the thermal contribution as a
function of radius and the high quality of the fits in the composite
model.

The overall profile of the remnant is asymmetrical, with a lobe
extending to the east. In order to determine whether this is emission
is powered strictly by an asymmetric relativistic wind from the pulsar
(cf.  Vela, Helfand et al 2001; B1509--58, Gaensler et al. 2002)
or represents the breakout of the hot gas into a region of low
surrounding density as is seen in some shell-type remnants, we
analyzed the spectrum of the emission extracted from the lobe region
defined above. Figure~5 displays the spectrum along with residuals
from a power law fit which fixed the column density at the value used
above for the remnant as a whole. The spectrum is well-characterized
by a steep power law with index $\Gamma = 2.88 \pm 0.06$ with no hint
of any thermal emission at low energies. The spectral index is the
steepest value we measured, consistent with its location farthest from
the pulsar (see Figure~6).

\section{Discussion}

In Figure~6, we illustrate the relative contributions of the thermal
and non-thermal components as a function of radius. The thermal shell
is clearly present. Note, however, that this thermal emission
represents less than~4\% of the observed (absorbed) flux of the
remnant in the $0.5-10$~keV band, explaining why it has proven so
elusive to date. Figure~7 displays the radial dependence of the shell
temperature and Neon abundance in the remnant shell -- both are
flat. In contrast, the power law spectral index $\Gamma$ monotonically
increases from $\Gamma = 1.9$ at the remnant center to $\Gamma = 2.8$
at the edge (and $\Gamma = 2.9$ in the eastern lobe). This latter
result is generally consistent with earlier work (Torii et al 2000; Slane
et al. 2002) although, owing to the much larger number of photons collected
and the spectral-spatial decomposition we have performed, the
measurements extend to a greater radius and differ in detail,
remaining flat between $2.2^{\prime}$ and $3.2^{\prime}$.

The widths of the annuli provide largely independent measurements of
the spectrum at each radius given the size of the PSF. Some
contamination is present from the inner, brighter rings to the outer
ones, however. Since the dominant power law is flatter in the central
regions, photons scattered to the outer regions bias the power law
measured there. We have determined the degree of contamination from
power law photons in each region by using the PSF binned in
$0.5^{\prime}$ annuli and the geometrical intersection of those annuli
with each extraction annulus (the minimal dependence of the PSF on
photon energy allows us to ignore this complicating factor). The
corrections to the measured power law slopes reported in Table~2 are
modest, albeit outside the formal errors. In the central
$0.5^{\prime}$ circle, the power law spectral index falls from 2.02 to
2.00, while in the first four annuli beyond this, the indices changes
as follows: 2.29 becomes 2.38, 2.45 becomes 2.50, 2.55 becomes 2.58,
and 2.73 becomes 2.78.  Thus, the rise in spectral slope is slightly
steeper than shown in Figure~7.  PSF contamination effects are
negligible for the thermal component since the count rate is so low.

The parameters we derive for the shell are in excellent agreement with
those inferred in the early work of Bocchino et al. (2001) and with
those found from a portion of the shell in the deep Chandra
observation of 3C58 (Slane et al. 2004): the temperature and Neon
abundance are identical at $kT=0.23$~keV and~3.2 solar,
respectively. While these previous authors have cited the Neon
overabundance as evidence that we are seeing emission from the ejecta,
a growing body of work (Cunha et al. 2006 and references therein)
suggests that the solar abundance of Neon has been systematically
underestimated by a factor of 2.5~to~3. In that our fits do not
require the statistically marginal enhancement of a factor of two in
the Mg abundance reported by Slane et al. (2004), we argue that the
provenance of the thermal emission remains an open question.

The total observed X-ray luminosity we derive for the shell component
is $5.9 \times 10^{32} d_{3.2}^2$~ergs s$^{-1}$, within the range
found in Bocchino et al. (2001). This is significantly below the upper
limit derived by Seward et al.  (2006) for the missing Crab Nebula
shell. The total mass is $\sim 0.77 M_{\odot}$ (for a mean molecular
weight of 0.6 which may not be appropriate if the bulk of the
radiating material is ejecta).

Evidence has been accumulating for some time that, despite the
apparently robust conclusions drawn from the historical records that
3C58 is coincident with SN1181 (Stephenson and Green 1999), the
remnant properties are inconsistent with such a young age.  Bietenholz
(2006) provides a comprehensive summary of the arguments against an
association, as well as a list of alternative scenarios. Most of the
evidence suggests a remnant age of $\sim 3000-5000$ years (e.g.,
Chevalier 2005). Our spatially resolved image of the shell adds
further to the case for an age $>820$~yrs.

Using the X-ray emitting mass we have derived and the analytic models
of PWN evolution from Chevalier (2005) we can set limits on the
apparent age of the remnant. From Eq.~27 of Chevalier (2005), we
can combine the observed supernova remnant radius of $2.8 \
d_{3.2}$~pc and the pulsar timing-derived parameter $\dot E = 2.7 \times
10^{37}$ erg s$^{-1}$ to obtain the age $t_3$ in units of 1000~yrs,

 $$t_3 \approx 1.8 \ M_{ej}^{0.4} \ E_{51}^{-0.2}$$

\noindent where $M_{ej}$ is the ejected mass in units of solar mass
$M_{\odot}$. If we make the extreme assumption that all of the
observed $0.77 M_{\odot}$ of X-ray-emitting gas is ejecta, and that it
represents some fraction $\leq 1$ of the total ejected mass of the
supernova, we find $t_3 \simgt 1.62 \ E_{51}^{-0.2}$.  Thus, the
pulsar must be at least twice the age of SN1181 for reasonable
explosion energies and, for a more plausible ejected mass of $\sim 5
M_{\odot}$, the age is 3400~yrs. Alternatively, if we assume
all of the observed gas has been swept up by the expanding PWN, from
Chevalier's (2005) eq. 28, we find $t = 5100$~yrs (cf. the pulsar's
characteristic age of 5390~yrs).

The pulsar's location with respect to the circular shell of thermal
X-ray emission provides a potential test that could resolve the age
controversy.  The location of the pulsar is marked with a crosshair in
Figure 8. It is apparent that this is not the center of the shell; a
circle best matching the observed thermal emission is centered
$27^{\prime\prime} \pm 5^{\prime\prime}$ west (and slightly north) of
the pulsar's location. For an age of 821 years, a distance of 3.2~kpc,
and spherical expansion (an assumption supported by the fact we see no
thermal emission in the eastern elongation of the remnant), this
requires a two-dimensional pulsar velocity of $500d_{3.2}$~km
s$^{-1}$. While not impossible for a young neutron star, it is worth
noting that if the object has the same tranverse velocity as the Crab
pulsar, the remnant age would be $3750d_{3.2}$~yrs, consistent with
other estimates. We further note that the implied trajectory for the
pulsar aligns to within a few degrees of the long axis of the nebula,
behavior commonly seen among PWNe. If Chandra survives for another
decade, it will be possible to resolve this issue directly, as the
implied proper motion of $0.5^{\prime\prime}$ for the high-velocity
scenario will be directly measurable.

\bigskip

\acknowledgements

This research is supported by NASA LTSA grant NAG~5-8063 to EVG and by
grant SAO~GO3-4026B to DJH. This research has also made use of data
obtained from the High Energy Astrophysics Science Archive Research
Center (HEASARC), provided by NASA's Goddard Space Flight Center.

\begin{deluxetable*}{lccccccc}
\tablewidth{0pt}
\tabletypesize{\scriptsize}
\tablecaption{Spectral Fits and Fluxes\label{spectra}}
\tablehead{
\colhead{Model}    & & & &\colhead {Annulus}  & &  & \\
\colhead{Parameter}& \colhead{(0\farcm0-0\farcm5)}& \colhead{(0\farcm5-1\farcm0)}& \colhead{(1\farcm0-1\farcm5)}& \colhead{(1\farcm5-2\farcm0)}& \colhead{(2\farcm0-2\farcm5)}& \colhead{(2\farcm5-3\farcm0)}& \colhead{(3\farcm0-3\farcm5)}
}
\startdata
$N_{\rm H}$ ($10^{21}$ cm$^{-2}$)& 4.16(4.09-4.24)      & 4.16(fixed)           & 4.16(fixed)     &  4.16(fixed)     & 4.16(fixed)     & 4.16(fixed)     & 4.16(fixed) \\
$\Gamma$ (spectral index)       & 2.02(2.01-2.04)       & 2.29(2.27-2.31)       & 2.45(2.43-2.48) &  2.55(2.51-2.59) & 2.73(2.67-2.78) & 2.81(2.70-2.92) & 2.93(2.75-3.12) \\
PL Flux\tablenotemark{a} pn     & $3.43\times 10^{-12}$ & $2.80\times 10^{-12}$ & $2.06\times 10^{-12}$ &  $1.50\times 10^{-12}$ & $8.40\times 10^{-13}$ & $3.86\times 10^{-13}$ & $2.05\times 10^{-13}$ \\
PL Flux\tablenotemark{a} MOS    & $4.14\times 10^{-12}$ & $3.15\times 10^{-12}$ & $2.38\times 10^{-12}$ &  $1.67\times 10^{-12}$ & $9.60\times 10^{-13}$ & $4.40\times 10^{-13}$ & $2.48\times 10^{-13}$ \\
$\chi^2$(DoF)                   &   404.01(442)         &   207.01(251)         &   138.71(196)   &    129.18(152)   &    80.28 (101)  &    36.98(66)    &    39.66(52) \\
\tableline
$k$T (keV)                      & 0.24(fixed)           & 0.24(0.20-0.29)       & 0.22(0.20-0.24) &  0.22(0.21-0.23) & 0.23(0.22-0.23) & 0.23(0.22-0.24) & 0.24(0.21-0.28) \\
Ne ([Ne/Ne$_{\sun}$])                                   & 3.0(fixed)            & 5.5(3.0-9.4)    & 4.3(3.6-4.8)    &  3.1(2.7-3.4)    & 3.0(2.8-3.3)    & 3.6(3.2-4.0)    & 3.1(2.2-4.4)    \\
kT Flux\tablenotemark{a} pn     & $ ... $               & $0.27\times 10^{-13}$ & $0.63\times 10^{-13}$ &  $1.16\times 10^{-13}$ & $1.39\times 10^{-13}$ &  $0.96\times 10^{-13}$ &  $0.26\times 10^{-13}$ \\
kT Flux\tablenotemark{a} MOS    & $ ... $               & $0.18\times 10^{-13}$ & $0.57\times 10^{-13}$ &  $1.16\times 10^{-13}$ & $1.63\times 10^{-13}$ &  $1.05\times 10^{-13}$ &  $0.26\times 10^{-13}$ \\
$\chi^2$(DoF)                   &    399.37(443)        &   373.58(414)         &   294.33(365)         &    345.91(319)         &   324.62(252)         &    140.76(164)         &    117.61(125)  \\
Total Flux\tablenotemark{a} pn  & $3.43\times 10^{-12}$ & $2.82\times 10^{-12}$ & $2.12\times 10^{-12}$ &  $1.62\times 10^{-12}$ & $9.76\times 10^{-13}$ &  $4.79\times 10^{-13}$ &  $2.34\times 10^{-13}$ \\
Total Flux\tablenotemark{a} MOS & $4.14\times 10^{-12}$ & $3.20\times 10^{-12}$ & $2.41\times 10^{-12}$ &  $1.78\times 10^{-12}$ & $1.12\times 10^{-12}$ &  $5.40\times 10^{-13}$ &  \phantom{0}$2.62\times 10^{-13}$
\enddata
\tablenotetext{a}{Absorbed flux in the 0.5--10 keV band in units of  (ergs cm$^{-2}$ s$^{-1}$).}
\tablecomments{Uncertainties are 90\% confidence for two interesting parameters.}
\end{deluxetable*}

\end{document}